\documentclass{jfm}
\usepackage{graphicx}
\usepackage{epstopdf, epsfig}

\shorttitle{Conditional statistics of supersonic turbulent boundary layer}
\shortauthor{X. Wu, J. Liang, Y. Zhao and M. Avila}

\title{Conditional statistics of supersonic turbulent boundary layer in the intermittent region}

\author{Xiaoshuai Wu\aff{1,2},
  Jianhan Liang\aff{1}
  \corresp{\email{jhleon@vip.sina.com}},
  Yuxin Zhao\aff{1}
 \and Marc  Avila\aff{2}}

\affiliation{\aff{1}Science and Technology on Scramjet Laboratory, National University of Defense Technology,
Changsha 410073, China
\aff{2}Center of Applied Space Technology and Microgravity, University of
Bremen, 28359 Bremen, Germany}

\begin{document}

\maketitle

\begin{abstract}
The turbulence statistics of the intermittent region in a supersonic turbulent boundary layer are studied by direct numerical simulation. Recently, \citet{kwon2016use} have shown that this intermittent behavior, consisting of the coexistence of turbulent and non-turbulent flows separated by an interface, could result in perils when the traditional Reynolds decomposition is used to analyze data. Here, we extend the study of turbulent/non-turbulent interfaces to the supersonic regime and report turbulence statistics conditioned to the interface. We show that the mathematical artefact of the traditional Reynolds/Favre decomposition contaminates the Reynolds stresses and thermal statistics. Our conditional statistics provide direct evidence that turbulent flow possesses positive Reynolds shear stress everywhere, whereas non-turbulent flow carries zero Reynolds shear stress. The thermal statistics of turbulent and non-turbulent flows exhibit opposite trends in the intermittent region, and the difference can be traced back to the variation of total temperature resulting from the Prandtl number.

\end{abstract}


\section{Introduction}\label{sec:introduction}

The zero-pressure gradient (ZPG) flat plate turbulent boundary layer (TBL) is an important canonical flow for experimental, theoretical and numerical studies. One notable feature of TBL, which is absent in internal flows, is the presence of a sharp but irregular boundary between turbulent and non-turbulent flow in the outer part, known as the turbulent/non-turbulent (T/NT) interface \citep{corrsin1943investigation}. Near the T/NT interface, the flow is intermittent in the sense that there is coexistence of large-scale irrotational and rotational fluid regions. An important implication of intermittency is that turbulence statistics should be treated with caution if an overall average of turbulent and non-turbulent regions is taken as the reference for fluctuations, namely within the scope of traditional Reynolds decomposition. \citet{jimenez2010turbulent} demonstrated that this mathematical artefact  may in fact veil the genuine physical mechanisms and thus conditional average should be used. Recently, \citet{kwon2016use} demonstrated the perils of the traditional Reynolds decomposition for analyzing turbulent structures and proposed a new decomposition to treat turbulent and non-turbulent regions separately. This ensures that non-turbulent regions in the flow do not contaminate the fluctuating velocity statistics.

\citet{morkovin1962effects} proposed that for moderate free-stream Mach numbers dilatation is small and hence differences with incompressible turbulence can be accounted for by mean fluid property variations. His hypothesis is the basis of van Driest transformation for the mean velocity profile \citep{van2003turbulent}, and is also the basis leading to the success of Morkovin's scaling  for Reynolds stress \citep{spina1994physics}. In recent years, the validity of Morkovin's hypothesis has been extensively checked from moderate supersonic flow to the hypersonic regime \citep{guarini2000direct, maeder2001direct, pirozzoli2004direct, duan2010direct, duan2011direct, lagha2011numerical}. Another important aspect of the study of compressible TBL is the strong Reynolds analogy (SRA) proposed by \citet{morkovin1962effects}, which relates the temperature fluctuations to velocity fluctuations and is widely used to extend incompressible turbulence models to compressible flows \citep{duan2010direct}. Lots of endeavors have been devoted to check the validity of SRA and several modifications have been proposed \citep{gaviglio1987reynolds, huang1995compressible, zhang2014generalized}. So far, the studies for supersonic TBL have largely employed the traditional Reynolds/Favre decomposition. Here we perform direct numerical simulation (DNS) of supersonic TBL and report turbulence statistics conditioned to T/NT interface location.
We show that the strong anti-correlation between velocity and temperature fluctuations assumed in the SRA breaks down as the T/NT interface is approached and this occurs much earlier than inferred from the traditional decomposition.

\section{Specification and validation of the DNS}\label{sec:specification and validation of the DNS}

\subsection{Numerical method and setup}

The supersonic boundary layer is simulated in a parallelepiped over a smooth no-slip wall. At the inflow boundary the laminar compressible boundary layer solution is imposed, whereas periodicity is prescribed in the span-wise direction and non-reflecting boundary conditions are used at the top boundary and outflow. Finite differences are employed to discretize the compressible Navier--Stokes equations in  conservative form on a Cartesian mesh. A WENO-SYMBO scheme \citep{martin2006bandwidth} and an eighth-order central scheme are utilized for the discretization of convective and viscous terms, respectively. Third-order TVD Runge--Kutta method is implemented for time advancement. Details of the method and implementation can be found in \citet{xin2006direct, li2010direct}.

The flow conditions correspond to a ZPG flat plate TBL at free-stream Mach number \textit{M}$_\infty$=2.87, and \textit{T}$_\infty$=101.99\textit{K}. The wall temperature is set to its nominal adiabatic value \textit{T}$_{aw}$/\textit{T}$_\infty$$=1+\textit{r}(\gamma-1)/2$\textit{M}$^2_\infty$, with the recovery factor \textit{r}=\Pran$^{1/3}$ and Prandtl number $\Pran=0.72$.
The Reynolds number based on the local momentum thickness and wall viscosity \citep{fernholz1977critical} of the laminar compressible boundary layer solution at the inlet  is \Rey$_{\delta2}$=$\rho$$_\infty$\textit{u}$_\infty$$\theta$/$\mu$$_w$=240. The blowing-and-suction approach of \citet{pirozzoli2004direct} is utilized to cause a rapid laminar-turbulent transition shortly downstream of the inlet, and a long enough computational domain is used to guarantee that the turbulent boundary layer is free from the tripping effect. The useful Reynolds number covers from \Rey$_{\delta2}$=1283 to \Rey$_{\delta2}$=1750, corresponding to \Rey$_{\tau}$=382 and \Rey$_{\tau}$=483, respectively. The computational domain is $L_x$$\times$$L_y$$\times$$L_z$=450$\delta$$_{in}$$\times$28$\delta$$_{in}$$\times$17.5$\delta$$_{in}$, $\delta$$_{in}$ being the boundary layer thickness at the inlet, with 2900$\times$120$\times$256 points in the stream-wise, wall-normal and span-wise directions, respectively. The height and width of the computational domain are chosen to be at least twice the largest 99\% boundary-layer thickness, which reaches $\delta$$_{99}$=7.7$\delta$$_{in}$ at \Rey$_{\delta2}$=1750. The grid spacing in wall units at \Rey$_{\delta2}$=1283 is $\Delta$\textit{x}$^+$$\times$$\Delta$\textit{y$_w$}$^+$$\times$$\Delta$\textit{z}$^+$=8.89$\times$0.89$\times$4.46, where the superscript + denotes quantities made dimensionless with the friction velocity \textit{u}$_\tau$=($\tau$$_w$/$\rho_w$)$^{1/2}$ and viscous length scale $\delta_\nu$=$\nu_w$/\textit{u}$_\tau$. In the following, the stream-wise, wall-normal and span-wise velocities are denoted as \textit{u}, \textit{v} and \textit{w}, respectively. We also follow the usual convention for the Reynolds ($f$=$\overline{f}$+$f'$) and Favre ($f$=$\widetilde{f}$+$f''$) decompositions, with $\widetilde{f}$=$\overline{{\rho}f}$/$\overline{\rho}$.

\begin{figure}
  \centerline{\includegraphics[width=12cm, height=5cm]{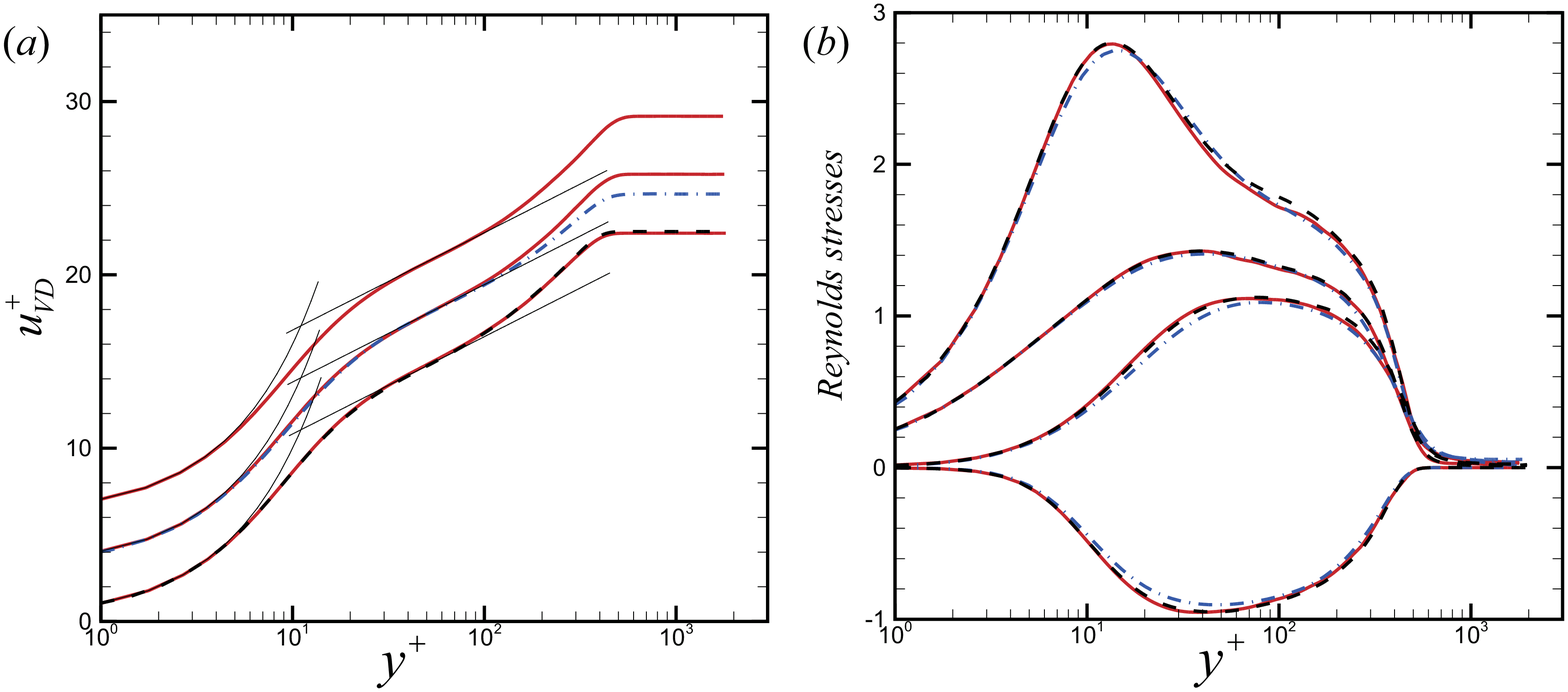}}
  \caption{(\textit{a}) Distribution of the van Driest transformed mean stream-wise velocity in wall units: (red) ---$\!$---, the present DNS at \Rey$_{\tau}$$=403, 445, 483$; (black) {-}{-}{-}{-}{-}, DNS by \citet{bernardini2011wall} at \Rey$_{\tau}$$=403$, \textit{M$_\infty$}=3; (blue) {-}{$\cdot$}{-}{$\cdot$}{-}{$\cdot$} DNS by \citet{jimenez2010turbulent} at \Rey$_{\tau}$$=445$. The linear and logarithmic regions are indicated by black thin lines, using \textit{u}$_{VD}^{+}$$=\textit{y}^{+}$, and \textit{u}$_{VD}^{+}$$=(1/0.41)\log(\textit{y}^{+})+5.2$, respectively. The profiles are shifted by \textit{u}$_{VD}^{+}$$=3$ for increasing \Rey$_{\tau}$. (\textit{b}) Density-scaled Reynolds stresses ($\overline{\rho}$/$\overline{\rho}$$_w$)$^{1/2}$\textit{u}$'$$_{\textit{rms}}^{+}$, ($\overline{\rho}$/$\overline{\rho}$$_w$)$^{1/2}$\textit{w}$'$$_{\textit{rms}}^{+}$, ($\overline{\rho}$/$\overline{\rho}$$_w$)$^{1/2}$\textit{v}$'$$_{\textit{rms}}^{+}$, and ($\overline{\rho}$/$\overline{\rho}$$_w$)$\overline{\textit{u$'$v$'$}}$$^{+}$ (from top to bottom): (red) ---$\!$---, the present DNS at \Rey$_{\tau}$$=452$; (black) {-}{-}{-}{-}{-},\citet{pirozzoli2011turbulence} at \Rey$_{\tau}$$=448$, \textit{M$_\infty$}=2; (blue) {-}{$\cdot$}{-}{$\cdot$}{-}{$\cdot$} DNS by \citet{jimenez2010turbulent} at \Rey$_{\tau}$$=445$.
  }
\label{fig:validation}
\end{figure}

\subsection{Validation of the database}

The mean stream-wise velocity profiles from the present DNS, scaled with the van Driest transformation (d\textit{u}$_{VD}$=($\overline{\rho}$/$\overline{\rho}$$_w$)$^{1/2}$d$\overline{\textit{u}}$),  are shown in figure~\ref{fig:validation}$(a)$ for stream-wise locations with \Rey$_{\tau}$$=403, 445, 483$. The transformed mean profiles agree well with the incompressible law-of-the-wall, showing a linear increase in the viscous sub-layer and a narrow logarithmic region in the overlap layer. The agreement with the data of \citet{bernardini2011wall} at \textit{M$_\infty$}=3 is excellent across the whole boundary layer. The collapse in the wake region confirms that the present TBL is fully developed and free from post-transitional effects. This is as expected because the Reynolds number at the tripping location is rather low and the flow responds very efficiently \citep{schlatter2012turbulent, eitel2014simulation}, making the effect of inflow conditions negligible after \Rey$_{\delta2}$=1283. Comparing with the incompressible data of \citet{jimenez2010turbulent}, we find an increase of wake strength with Mach number for the same \Rey$_{\tau}$, which is consistent with previous studies \citep{bernardini2011wall, zhang2012mach}.

During the simulation data was collected at the stream-wise location with \Rey$_{\tau}$=452, including 38750 samples in the span-wise-wall-normal (transverse) plane $(y,z)$ for a period of 127$\delta$$_{99}$/\textit{u}$_\infty$, which is long enough to obtain converged statistics. Figure~\ref{fig:validation}$(b)$ depicts the density-scaled Reynolds stresses distributions from this database in inner scaling, and includes the closest available incompressible Reynolds number from \citet{jimenez2010turbulent}, as well as the compressible DNS by \citet{pirozzoli2011turbulence}. An excellent collapse is achieved throughout the whole boundary layer. The valuable information here is that an asymptotic equilibrium turbulent state can be achieved after the initial tripping effect becomes negligible, despite employing different turbulence generation methods from \citet{jimenez2010turbulent} and \citet{pirozzoli2011turbulence}, who used the recycling procedure. In all, the good agreement demonstrates that a generic supersonic ZPG TBL is faithfully simulated and the data sets are reliable.

\begin{figure}
  \centerline{\includegraphics[width=12cm, height=5cm]{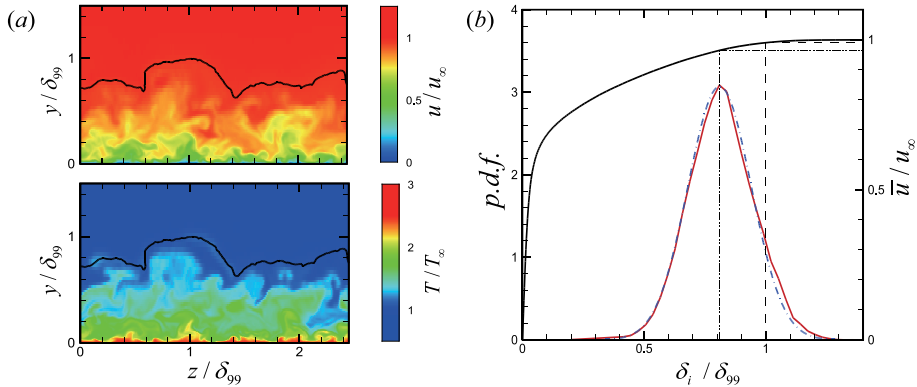}}
  \caption{($a$) Colormaps of instantaneous stream-wise velocity (top) and temperature (bottom) at  \Rey$_{\tau}=452$. The T/NT interface is shown as a black line. ($b$) Properties of T/NT interface: (red) ---$\!$---, p.d.f. of the interface location $\delta_i$ scaled by $\delta_{99}$; (blue) {-}{$\cdot$}{-}{$\cdot$}{-}{$\cdot$}, normal distribution with mean $\mu$=$0.81\delta_{99}$ and standard deviation $\sigma$$=0.13\delta_{99}$; (black) ---$\!$---, temporal mean velocity profile $\overline{u}$/\textit{u}$_\infty$. The mean T/NT interface and boundary layer edge locations are denoted by dash-dot-dotted and dashed lines, respectively. }
\label{fig:TNTI}
\end{figure}

\section{Results}\label{sec:results}

\subsection{Detection of the T/NT interface}

For the detection of the T/NT interface, rather than employing a criterion based on the local turbulent kinetic energy \citep{chauhan2014turbulent}, we assess the interface location by using a constant stream-wise velocity magnitude of $97$\%\textit{u}$_\infty$ \citep{de2017interfaces}. This criterion is able to distinguish the turbulent flow in light of the existence of uniform momentum zones in turbulent boundary layers, as shown by \citet{de2017interfaces}. In addition, the instantaneous interface is intensely folded, and multiple T/NT interfaces could be met as one approaches the wall from the free stream. Following \citet{de2017interfaces}, we consider the upper envelope of the interface as the nominal T/NT interface location in order to ensure that the flow above the detected interface is non-turbulent. An example of the described detection technique, applied to an instantaneous flow field at \Rey$_{\tau}=452$, is shown in figure~\ref{fig:TNTI}$(a)$. By examining the streamwise velocity and temperature fields, it is seen that the turbulent region is indeed well encapsulated by the T/NT interface.

Figure~\ref{fig:TNTI}$(b)$ illustrates the probability density function ({p.d.f.}) of the instantaneous T/NT interface height $\delta_i$, scaled by the local 99\% boundary layer thickness $\delta_{99}$. The p.d.f. exhibits a nearly normal distribution with mean $\mu$=$0.81\delta_{99}$ and standard deviation $\sigma$$=$0.13$\delta_{99}$. For an incompressible boundary layer, \citet{chauhan2014turbulent} found that the {p.d.f.} is also approximately normal with  $\mu$=$0.67\delta$ and $\sigma$$=0.11\delta$, where $\delta$ is a boundary layer thickness determined by fitting a composite velocity profile \citep{chauhan2009criteria}. Considering that $\delta$ is approximately 25\% greater that the 99\% boundary layer thickness \citep{kwon2016use}, our data yield $\mu$=$0.65\delta$ and $\sigma$$=0.11\delta$. Noting the higher Reynolds number in \citet{chauhan2014turbulent}, the interesting finding here is that the p.d.f. of the interface location seems neither affected by Reynolds number nor influenced by compressibility effects. The good agreement not only confirms the proper implementation of detecting the T/NT interface, but also demonstrates that $\delta_{99}$ works as a robust scaling for the interface location.

\begin{figure}
  \centerline{\includegraphics[width=12cm, height=5cm]{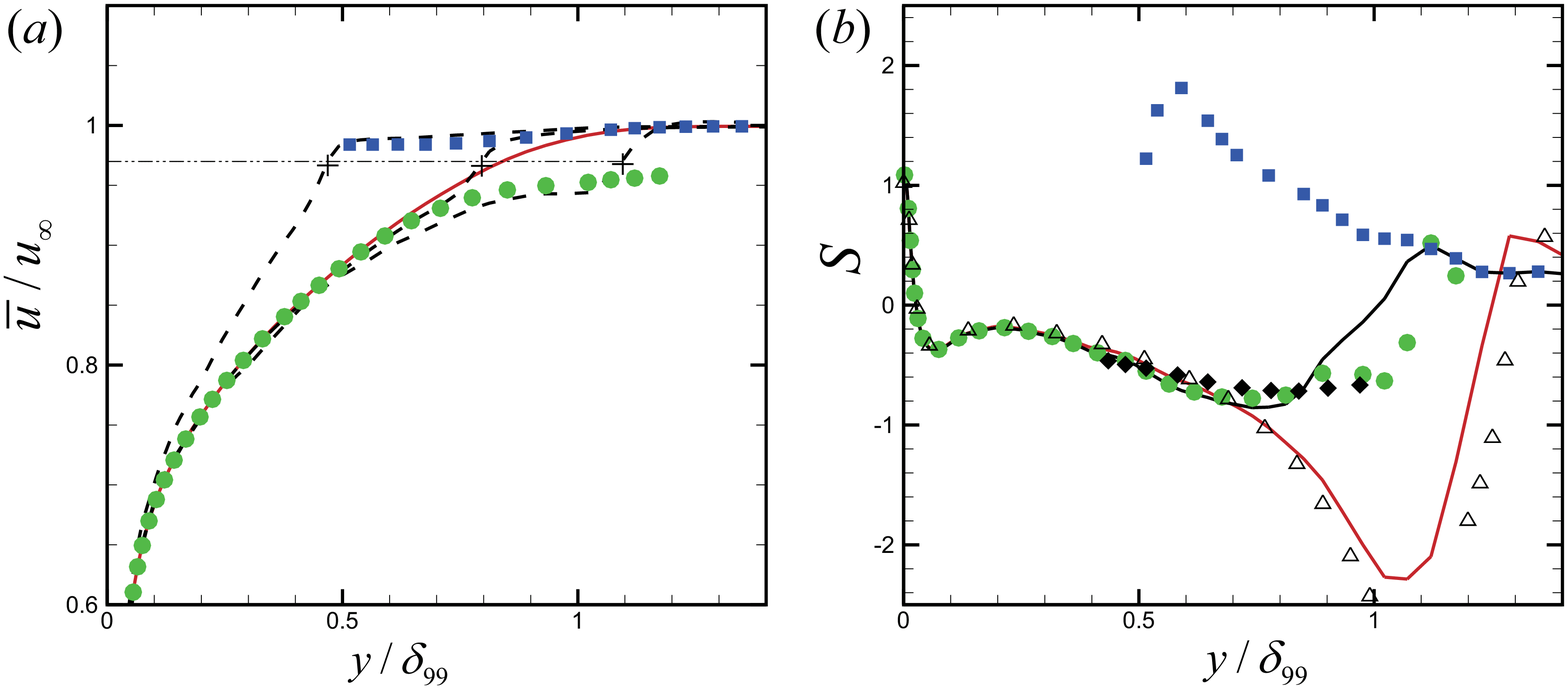}}
  \caption{Profiles of conditional statistics at \Rey$_{\tau}$$=452$. (\textit{a}) Conditional mean velocity profiles (black dashed lines) for different T/NT interface locations $\delta_{i}$/$\delta_{99}$=$0.47, 0.80, 1.10$ (from left to right), with the bin being $0.03\delta_{99}$. The red solid line is the (traditional) temporal mean velocity, $\overline{u}$/\textit{u}$_\infty$. The conditional turbulent mean and non-turbulent mean are shown as green bullets and blue squares, respectively. The nominal T/NT interface locations are marked by $``$+$"$ signs. The thin dash-dot-dotted line denotes the criterion of 97\%\textit{u}$_\infty$. (\textit{b}) Skewness of streamwise velocity fluctuations determined under the traditional Reynolds decomposition (red solid line) and the new decomposition (black solid line); (green) $\bullet$, the conditional turbulent skewness with fluctuations determined under the new decomposition;  (blue) $\blacksquare$, the conditional non-turbulent counterpart; $\vartriangle$, the compressible DNS by \citet{pirozzoli2011turbulence} at \Rey$_{\tau}$$=447$, \textit{M$_\infty$}$=2$; $\blacklozenge$, channel flow by \citet{gunther1998turbulent} at \Rey$_{\tau}$$=300$.
  }
\label{fig:meanvelocity}
\end{figure}

\subsection{Mean profiles}

We calculate conditionally averaged profiles by using the new decomposition of \citet{kwon2016use}. Here the averaging is conditioned to a small bin of T/NT interface locations, where the midpoint is considered as the nominal interface location. Then for a particular interface location, all the relevant instantaneous profiles from all the recorded samples are collected and averaged. The conditional mean profile is thus a function of the wall-normal coordinate and the instantaneous T/NT interface location, and the fluctuating component is obtained by subtracting the conditional mean from the instantaneous profile at the same stream-wise location. Because of the separate treatment for turbulent and non-turbulent flow regions, this decomposition removes the influence of T/NT interface movements.
Since our data is collected in the transverse plane and contains enough velocity vector fields for reasonable convergence, no averaging in the stream-wise domain is applied, which avoids a bias in the velocity fluctuations \citep{kwon2016use}.

As the counterpart of the traditional Reynolds decomposition, we use $\overline{f}$$|$$_{\delta_i}$ to denote the conditionally averaged mean of the instantaneous variable profiles \textit{f} when the interface is at the specific location $\delta_i$, and $f'$$|$$_{\delta_i}$ refers to the fluctuations under the new decomposition with the relation $f'$$|$$_{\delta_i}$=$f$-$\overline{f}$$|$$_{\delta_i}$, which means that the fluctuation at certain wall-normal position is also a function of the instantaneous T/NT interface location. The same convention is used for the Favre representation, being $f$=$\widetilde{f}$$|$$_{\delta_i}$+$f''$$|$$_{\delta_i}$ and $\widetilde{f}$$|$$_{\delta_i}$=$\overline{{\rho}f}$$|$$_{\delta_i}$/$\overline{\rho}$$|$$_{\delta_i}$. Besides the mean profiles conditioned to the specific nominal T/NT interface location, one may also be curious about the statistics conditioned to either kind of fluid, that is excluding the influence of intermittency. We use $\langle$$\overline{f}$$|$\textit{y}$<$$\delta_i$$\rangle$ to denote the conditional turbulent mean of the instantaneous profiles ${f}$ when the respective local wall-normal position is below the TNTI location, namely in the turbulent region, and the opposite is true for the conditional non-turbulent mean $\langle$$\overline{f}$$|$\textit{y}$>$$\delta_i$$\rangle$. For high order conditional statistics, the fluctuations are firstly computed with the new decomposition and then conditioned to either kind of fluid to obtain the conditional turbulent and non-turbulent values.

Figure~\ref{fig:meanvelocity}$(a)$ depicts the conditionally averaged streamwise velocity at various $\delta_i$. As expected, the conditional mean profile (dashed line) deviates substantially from the temporal mean (solid line) in the intermittent region, whereas the near-wall region is insensitive to the chosen decomposition method. At the nominal T/NT interface locations (marked with a cross for each profile) steep changes in velocity are found, in agreement with previous studies \citep{chauhan2014turbulent, kwon2016use}. Also included in figure~\ref{fig:meanvelocity}$(a)$ are the conditional turbulent and non-turbulent mean velocity profiles shown as symbols. It can be seen that the conditional turbulent mean velocity follows closely the temporal mean in the near-wall region, then deviates from it in the intermittent region with a lower magnitude as a consequence of the exclusion of high-speed non-turbulent flow. The conditional non-turbulent mean velocity achieves higher value than the temporal mean, and exhibits a minor decrease when approaching the wall from the free stream, which is a sign of deceleration by the surrounding low-speed turbulent regions. The present observations are consistent with the description by \citet{Kovasznay1970} that the potential fluid in boundary layers is much closer to the free stream than the temporal average.

Regarding the different properties of turbulent and non-turbulent flows, an intuitive impression can be gained by inspecting the skewness of the stream-wise velocity fluctuations shown in figure~\ref{fig:meanvelocity}$(b)$. By comparing the skewness obtained with the traditional (red line) and with the new (black line) Reynolds decomposition, it appears that in the intermittent region a large portion of the unconditional skewness can be attributed to the mathematical artefact produced by T/NT interface movements. Also included in figure~\ref{fig:meanvelocity}$(b)$ is the conditional turbulent skewness $\langle$\textit{S}$($$u'$$|$$_{\delta_i}$$)$$|$\textit{y}$<$$\delta_i$$\rangle$, as well as the non-turbulent one $\langle$\textit{S}$($$u'$$|$$_{\delta_i}$$)$$|$\textit{y}$>$$\delta_i$$\rangle$. It can be seen that in the intermittent region the turbulent flow shares some similarity with turbulent channel flow \citep{gunther1998turbulent}, given that ${\delta_{99}}$ works as an analogue of the channel half-width \citep{jimenez2010turbulent}, whereas the non-turbulent flow appears as high-speed inrush.

\begin{figure}
  \centerline{\includegraphics[width=12cm, height=5cm]{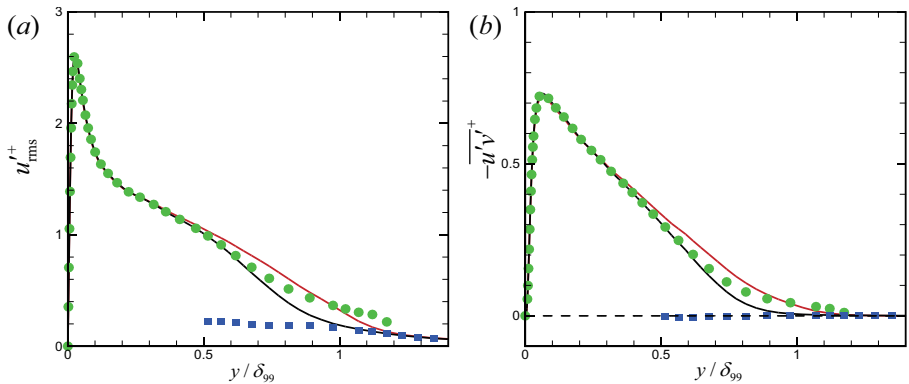}}
  \caption{ Distribution of (\textit{a}) the Reynolds normal stress \textit{u}$'$$_{\textit{rms}}^{+}$ and (\textit{b}) shear stress $-$$\overline{\textit{u$'$v$'$}}$$^{+}$ with velocity fluctuations determined under the traditional Reynolds decomposition (red solid line) and the new decomposition (black solid line); (green) $\bullet$, the conditional turbulent mean with velocity fluctuations under the new decomposition; (blue) $\blacksquare$, the conditional non-turbulent counterpart.
  }
\label{fig:Reynoldsstress}
\end{figure}

\subsection{Reynolds stresses}

Figure~\ref{fig:Reynoldsstress}$(a)$ shows the Reynolds normal stress computed with the traditional Reynolds decomposition (red line) and with the new decomposition (black line). A considerable portion of the traditional Reynolds normal stress in the intermittent region (approximately \textit{y}$>$0.5$\delta_{99}$) can be attributed to T/NT interface movements. This is consistent with \citet{kwon2016use} in light of the close relationship between turbulence intensities and power spectral densities. The near-wall region remains unaffected by the new decomposition.

By conditioning to either kind of fluid, the influence of intermittency can be further excluded to explore the intrinsic Reynolds stresses. Figure~\ref{fig:Reynoldsstress}$(a)$ also includes the Reynolds normal stress conditioned to the turbulent and non-turbulent regions. The conditional turbulent Reynolds normal stress achieves higher magnitude than the temporal mean computed under the new decomposition, as compensation for intermittency. The conditional non-turbulent Reynolds normal stress is found to be non-zero and is roughly constant across the intermittent region. In addition, the non-turbulent flow also possesses non-zero fluctuation magnitudes for the transverse components (not shown here). This can be explained as that the non-turbulent flow in the intermittent region, coming from the free stream, interacts with the surrounding turbulent flow and is either accelerated or decelerated. This is in line with the conditional mean velocity profiles shown in figure~\ref{fig:meanvelocity}$(a)$.

The fact that all three velocity components possess non-zero fluctuations in non-turbulent flow does not imply that Reynolds shear stresses are carried by non-turbulent flow. Figure~\ref{fig:Reynoldsstress}$(b)$ shows that the conditional non-turbulent Reynolds shear stress has zero magnitude across the intermittent layer. Note that the conditional shear stress is positive everywhere in the turbulent flow. Overall, the picture emerging from the conditional statistics is consistent with the physical mechanisms governing turbulent and non-turbulent flows, and it is in this sense that the conditional average is needed in the intermittent region. In addition, our observations are consistent with the statements of \citet{jimenez2010turbulent}, despite adopting different conditional approaches.

\begin{figure}
  \centerline{\includegraphics[width=12cm, height=5cm]{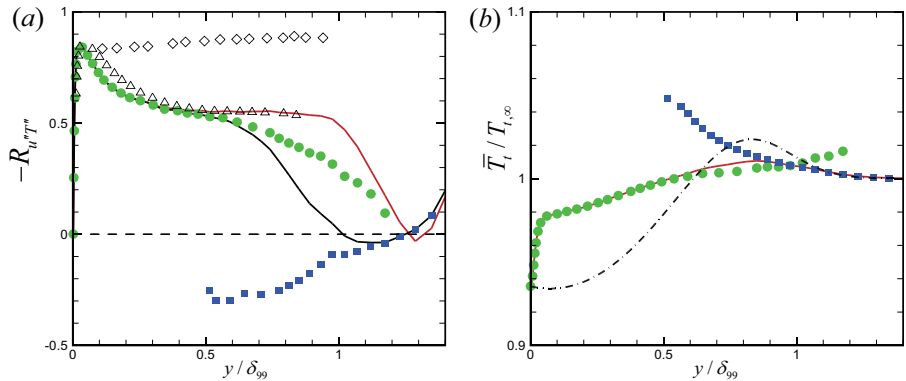}}
  \caption{ (\textit{a}) Distributions of velocity-temperature correlation with variable fluctuations determined under the traditional Favre decomposition (red solid line) and the new decomposition (black solid line); (green) $\bullet$, the conditional turbulent mean with fluctuations determined with the new decomposition; (blue) $\blacksquare$, the conditional non-turbulent counterpart; $\vartriangle$ the DNS by \citet{pirozzoli2011turbulence} at \Rey$_{\tau}$$=251$, \textit{M$_\infty$}=2; $\lozenge$, the compressible experiments by \citet{debieve1983etude} at \Rey$_{\delta2}$$=2720$, \textit{M$_\infty$}=3. (\textit{b}) Distribution of the total temperature normalized by the free-stream value in outer scaling: (red) ---$\!$---, the traditional temporal mean; (green) $\bullet$, the conditional turbulent mean; (blue) $\blacksquare$, the conditional non-turbulent mean; (black) {-}{$\cdot$}{-}{$\cdot$}{-}{$\cdot$} the total temperature of supersonic laminar boundary layer at \textit{M$_\infty$}=2.87 with $\Pran=0.72$ and adiabatic wall condition.
  }
\label{fig:SRA}
\end{figure}

\subsection{Thermal statistics}

The strong Reynolds analogy put forward by \citet{morkovin1962effects} implies in its strict form that velocity and temperature fluctuations are perfectly anti-correlated, with one form being \textit{R}$_{u''T''}$$=-1$. The temperature-velocity correlation $-$\textit{R}$_{u''T''}$ computed under the traditional Favre decomposition is depicted in figure~\ref{fig:SRA}$(a)$ as a red line. It attains its maximum in the buffer layer and maintains approximately 0.6 throughout most of the boundary layer, which is consistent with the literature \citep[e.g.][]{pirozzoli2011turbulence}. Although the velocity and temperature fluctuations are not perfectly anti-correlated \citep{maeder2001direct, pirozzoli2004direct}, the high anti-correlation indicates that large-scale eddies moving away from the wall contain warmer, lower-speed fluid than the average value found at the corresponding location \citep{spina1994physics}.

The contamination of T/NT interface movements can be distinguished by examining the correlation $-\textit{R}$$_{\textit{u$''$}|{\delta_i}\textit{T$ ''$}|{\delta_i}}$ (black line in figure~\ref{fig:SRA}$(a)$), which is a temporal mean computed from the fluctuations {\textit{u$''$}$|$$_{\delta_i}$ and {\textit{T$''$}$|$$_{\delta_i}$ under the new decomposition. The intrinsic velocity-temperature correlation in the turbulent and non-turbulent regions can be obtained from the corresponding conditional SRA. Considering the relative passive role of temperature, one may expect that the turbulent flow in the outer region ought to behave similarly with that in the inner region, hence achieving comparable values of the velocity-temperature correlation. However, the conditional turbulent SRA  $<$$-\textit{R}$$_{\textit{u$''$}|{\delta_i}\textit{T$ ''$}|{\delta_i}}$$|$\textit{y}$<\delta_i$$>$ exhibits a rapid decrease from the buffer layer even after compensation for intermittency, and the commonly accepted velocity-temperature anti-correlation is overestimated. As expected the free stream shows a negative correlation between velocity and temperature fluctuations (not shown here), but the interesting finding here is that, in the intermittent region, velocity and temperature fluctuations of the non-turbulent flow are positively correlated.

One possible reason for the behavior in the intermittent region is that interaction happens between the turbulent flow and non-turbulent region, during which the thermal and velocity fields are nonlinearly coupled. Some insights can be gained by inspecting the total temperature across the whole boundary layer, shown in figure~\ref{fig:SRA}$(b)$. The Prandtl number sets the ratio of the diffusion coefficients of momentum and thermal energy and determines the manner in which energy is redistributed within the whole layer. With adiabatic thermal boundary conditions and Prandtl number less than unity, the Reynolds averaged mean total temperature exhibits a noticeable accumulation of energy in the outer part of turbulent boundary layer, consistent with the observation of \citet{pirozzoli2004direct}. Here the nonlinear coupling between thermal and velocity fields leads to different behaviors for turbulent and non-turbulent flows. Figure~\ref{fig:SRA}$(b)$ shows that the turbulent flow displays a continuously increasing total temperature when departing from the wall, whereas the non-turbulent flow exhibits the opposite trend. Therefore, the positive correlation between velocity and temperature fluctuations for non-turbulent flow can be attributed to the increase of total temperature in the intermittent region. It is also found that in the intermittent region the non-turbulent flow does not behave as purely laminar flow. This is sensible because the non-turbulent flow experiences interactions with surrounding turbulent flows.

\section{Conclusions}\label{sec:conclusions}

In the present study, a generic supersonic ZPG TBL is computed for the study of turbulence statistics. An asymptotic equilibrium turbulent state is achieved after the tripping effect becomes negligible. It is demonstrated that the T/NT interface remains largely unaffected by compressibility, with 99\% boundary layer thickness working as a robust scaling for the interface height. The detection of T/NT interface allows a separate treatment of the turbulent and non-turbulent flow regions by using a recently proposed decomposition \citep{kwon2016use}. From this statistics conditioned to either kind of flow can be performed. In agreement with previous observations, we demonstrate the mathematical artefact incurred by T/NT interface movements in traditional statistics. By resorting to conditional average, a description consistent with the physical mechanism can be provided. Further we provide direct evidence that the turbulent flow possesses positive Reynolds shear stress everywhere, whereas the non-turbulent flow carries zero shear stress despite exhibiting non-zero fluctuation intensities in all three velocity components.

The SRA stems from the similarity between the momentum and energy equations. However, we found that caution should be taken when using SRA to characterize the flow, especially in the intermittent region, where the fluctuation of the T/NT interface is prominent. The results demonstrate that in the intermittent region the T/NT interface as well as the underlying intermittency influence the traditional quantification of SRA. The turbulent and non-turbulent flows feature opposite trends in the velocity-temperature correlation, and the difference can be traced back to the variation of total temperature caused by the Prandtl number under the adiabatic wall boundary condition. A remaining challenge is the interpretation of the T/NT interface or intermittency in SRA. Here it is difficult reconciling the mathematical foundation with the physical mechanisms, especially in view of the success of a generalized Reynolds analogy \citep{zhang2014generalized} even under the traditional Reynolds/Favre decomposition.

\section*{Acknowledgements}\label{sec:acknowledgements}

This research work is supported by National Natural Science Foundation of China under grant 11472304, and the Priority Programme SPP 1881 Turbulent Superstructures of the Deutsche Forschungsgemeinschaft. X.W. acknowledges financial support from China Scholarship Council under grant no. 201603170296.
\bibliographystyle{jfm}
\bibliography{jfm-instructions}

\end{document}